\documentclass[numreferences]{kluwer}

\usepackage{graphicx}

\begin{document}
\begin{article}
\begin{opening}

\title{Exponential distribution of long heart beat intervals\\ during
  atrial fibrillation and their relevance\\ for white noise behaviour in
  power spectrum}

\author{Thomas \surname{Hennig}\email{thomas.hennig@tu-ilmenau.de}} 

\author{Philipp \surname{Maass}\email{philipp.maass@tu-ilmenau.de}}
\institute{Institut f\"ur Physik, Technische Universit\"at Ilmenau,
  98684 Ilmenau, Germany}

\author{Junichiro \surname{Hayano}\email{hayano@med.nagoya-cu.ac.jp}}
\institute{Third Department of Internal Medicine, Nagoya City
  University Medical School, Nagoya 467, Japan}

\author{Stefan  \surname{Heinrichs}\email{stefan.heinrichs@uni-konstanz.de}} 
\institute{Fachbereich Physik, Universit\"at Konstanz, 78457
  Konstanz, Germany}

\runningtitle{Exponential distribution of long heart beat intervals
  during AF}

\runningauthor{Th.\ Hennig, Ph.\ Maass, J.\ Hayano, St.\ Heinrichs}


\date{10 March 2006}

\begin{abstract}
  The statistical properties of heart beat intervals of 130 long-term
  surface electrocardiogram recordings during atrial fibrillation (AF) are
  investigated.  We find that the distribution of interbeat intervals
  exhibits a characteristic exponential tail, which is absent during
  sinus rhythm, as tested in a corresponding control study with 72
  healthy persons. The rate $\gamma$ of the exponential decay lies in
  the range 3-12~Hz and shows diurnal variations. It equals, up to
  statistical uncertainties, the level of the previously uncovered
  white noise part in the power spectrum, which is also characteristic
  for AF. The overall statistical features can be described by
  decomposing the intervals into two statistically independent times,
  where the first one is associated with a correlated process with
  $1/f$ noise characteristics, while the second one belongs to an
  uncorrelated process and is responsible for the exponential tail. It
  is suggested to use $\gamma$ as a further parameter for a better
  classification of AF and for the medical diagnosis. The relevance of
  the findings with respect to a general understanding of AF is
  pointed out.
 \end{abstract}
 

\end{opening}

\section{Introduction}

Atrial fibrillation (AF) is the most common arrhythmia of the heart
and leads to an impairment of physical ability and to a reduction of
quality of life (for a review, see e.g.\ \cite{Nattel:2002}).
Connected with these degradations is an increased risk of
thromboembolic complications, of strokes \cite{Wolf/etal:1991} and of
mortality \cite{Benjamin/etal:1998}. The prevalence of AF in the
industrialised countries currently is about 1\% with respect to the
total population and gives rise to tremendous costs for diagnosis and
therapy. The risk to develop AF increases with age: For persons older
than 40 years 2.3\% are affected, for persons older than 60 years
5.9\%, and for persons older than 80 years 8\% \cite{Krahn/etal:1995}.
Because of the rise of life expectancy in the industrialised
countries, the number of persons with AF is likely to increase
further.

To support the choice between currently employed therapeutical
measures, such as pharmaceutical treatment with antiarrhythmica,
AV-nodal ablation or modulation, pulmonary vein or isthmus ablation,
and cardioversion, it is necessary to develop a good classification of
AF. The size of the left atrium is to date the main parameter to
support diagnosis and decisions for treatment options
\cite{Hoglund/etal:1985,Verhorst/etal:1985}, although its role in
predicting outcome after a cardioversion is discussed controversially
\cite{Manabe/etal:1997,Holm:1998}.  Further classifications are based
on the averaging of P waves \cite{Yamada/etal:1999,Raitt/etal:2000},
on the amplitude of the fibrillatory waves in the surface
electrocardiogram (ECG) \cite{Peter/etal:1966}, or on the morphology
of the intra-atrial ECG \cite{Wells/etal:1978}. The number and degree
of spatial uniformity of wave fronts was used for a classification
into 3 different AF types \cite{Konings/etal:1994}. In recent studies
short cycle lengths of fibrillatory waves were found to correlate with
the risk, that paroxysmal AF changes into persistent AF
\cite{Asano/etal,Boahene/etal:1990} and that arrhythmica fail to
induce conversion to the sinus rhythm
\cite{Stambler/etal:1997,Bollmann/etal:2002}.

All these established methods use information from intra-atrial
signals (or atrial contributions in the surface ECG) to classify AF.
However, also ventricular beats show characteristic signatures of AF
\cite{Moody/etal:1983,Pinciroli/etal:1986,Hayano/etal:1997}, parts of
them have been suggested for the detection of AF
\cite{Moody/etal:1983,Pinciroli:1993}.  In this work we will show that
the distribution of RR intervals during AF exhibits a characteristic
exponential tail with a decay rate $\gamma$ in the range 3-12~Hz. We
show that this rate can be determined also in the power spectrum. Our
analysis suggests to use $\gamma$ as an additional classification
parameter for AF.

\section{Characteristic features of atrial fibrillation\\ in RR interval
  distributions}

\begin{figure}[h!]
  \vspace*{3em}
  \includegraphics[width=12cm,clip=,]{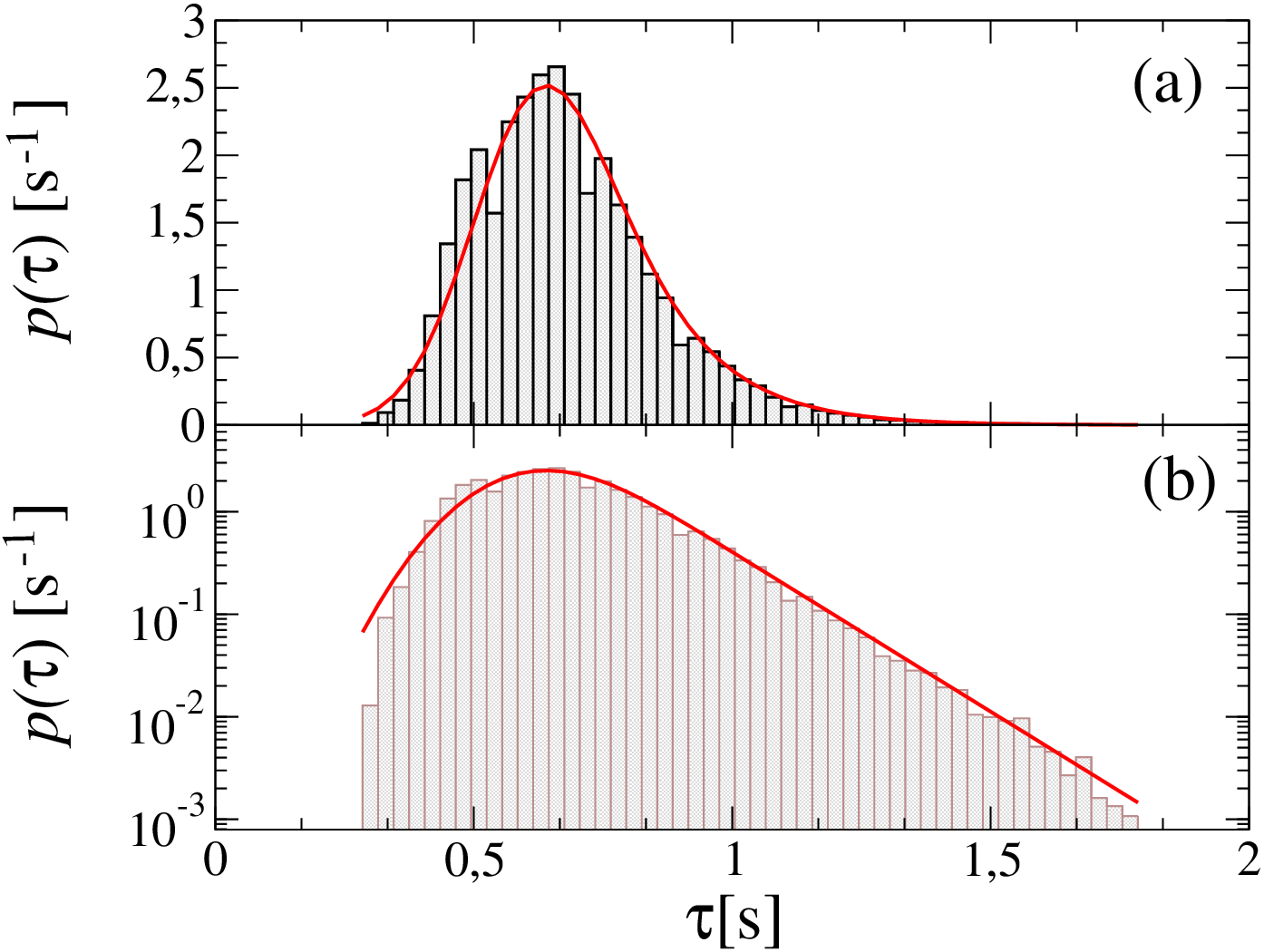}\\[2ex]
  \includegraphics[width=12cm,clip=,]{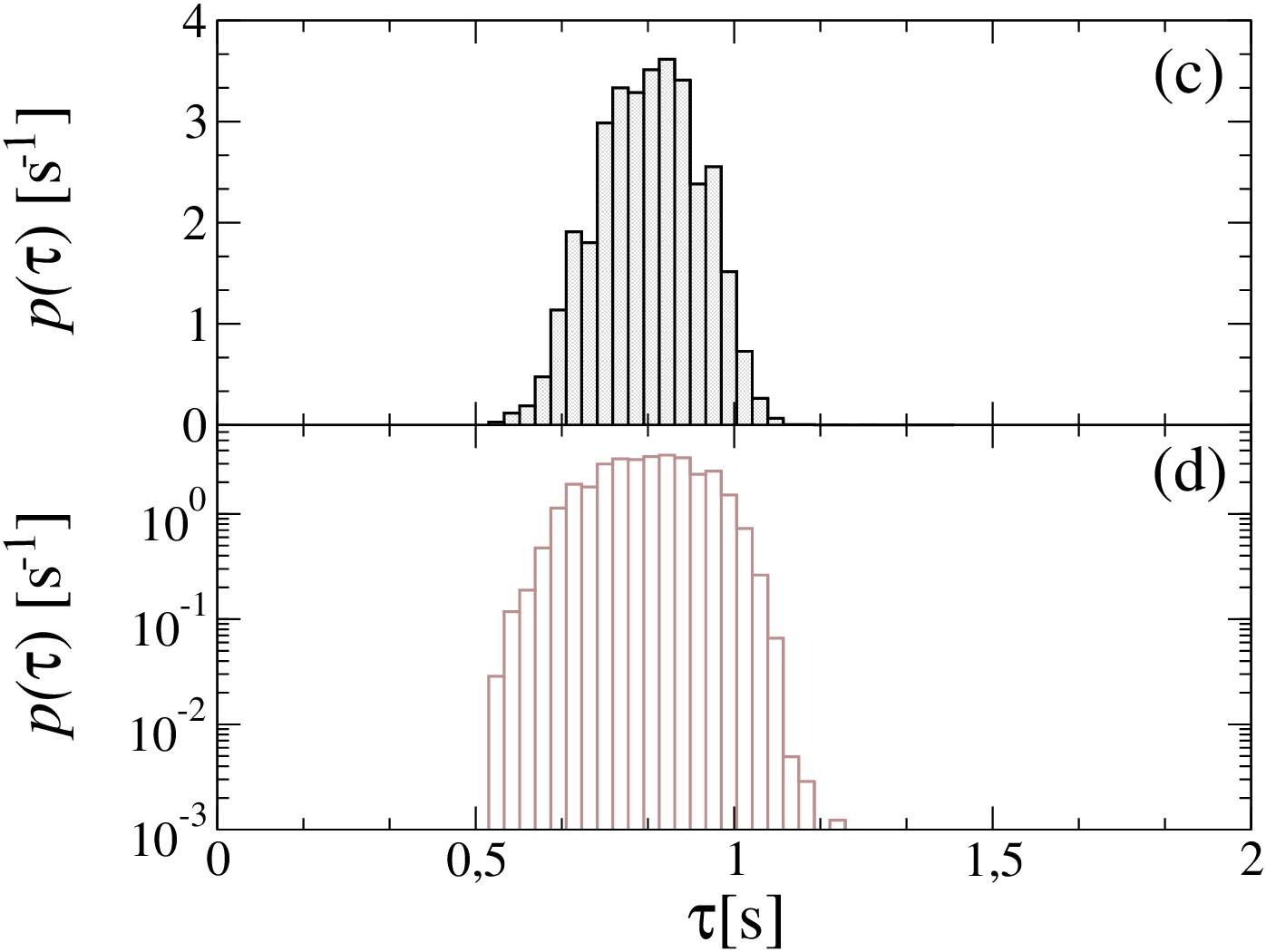}\\[0.5ex]
\caption{Probability density $p(\tau)$ of RR intervals $\tau$ 
  for one representative AF patient on {\it (a)} linear and {\it (b)}
  semi-logarithmic scale. In the calculation a fixed bin width of
  0.03~s\ was used.  The lines mark the fit to eq.~(\ref{eq:pfit})
  with the exponential decay for large $\tau$ (cf.\ 
  eq.~(\ref{eq:decay})). Corresponding plots on {\it (c)} linear and
  {\it (d)} semi-logarithmic scale of one representative healthy
  person show no signature of an exponential decay.}
\label{fig:density}
\end{figure}

Our study is based on RR sequences $\tau_n$ of 130 AF patients, which
typically comprise 10$^5$ beats (corresponding to 24h ECG recordings
\cite{Hayano:1997}). The probability density $p(\tau)$ of RR
intervals for one representative patient is shown in
Figs.~\ref{fig:density}a,b. We find that for large $\tau$, the decay
of $p(\tau)$ can be well fitted to an exponential,
\begin{equation}
p(\tau)\sim p_{\infty}\exp(-\gamma\tau)\,,
\label{eq:decay}
\end{equation}
with decay rate $\gamma$ and amplitude factor $p_\infty$. This
behaviour becomes particularly clear in the semi-logarithmic
representation in Fig.~\ref{fig:density}b.

We find that eq.~(\ref{eq:decay}) is a generic feature for AF
patients. To quantify the significance of (\ref{eq:decay}) we perform
a Kolmogorov-Smirnov test \cite{Honerkamp:1994} for the RR intervals
in the tail of $p(\tau)$. For sufficient statistical meaning, we
require the tail region to encompass at least 0.3~s and at
least 2\% of the RR intervals. We find that 106 of the 130 AF patients
(81.5\%) pass the Kolmogorov-Smirnov test with a standard significance
level of 10\%. Even better values are obtained by a more sophisticated
procedure for identifying the parameters $\gamma$ and $p_\infty$ from
an optimal fitting region for eq.~(\ref{eq:decay}) as described below.

One can expect that eq.~(\ref{eq:decay}) is also specific for atrial
fibrillation, since (\ref{eq:decay}) it is not known to be a
particular feature of healthy persons. To perform a countercheck, we
analyse RR sequences of 72 healthy persons, which were taken from the
PhysioNet database \cite{physionet}. The result for one
representative person is displayed in Figs.~\ref{fig:density}c,d.
Clearly, no exponential tail can be identified. Only 6 of the 72
healthy persons (8.3\%) would be accepted according to the
Kolmogorov-Smirnov test as specified above.

\begin{figure}[h!]
  \vspace*{1em}
  \includegraphics[width=10cm,clip=,]{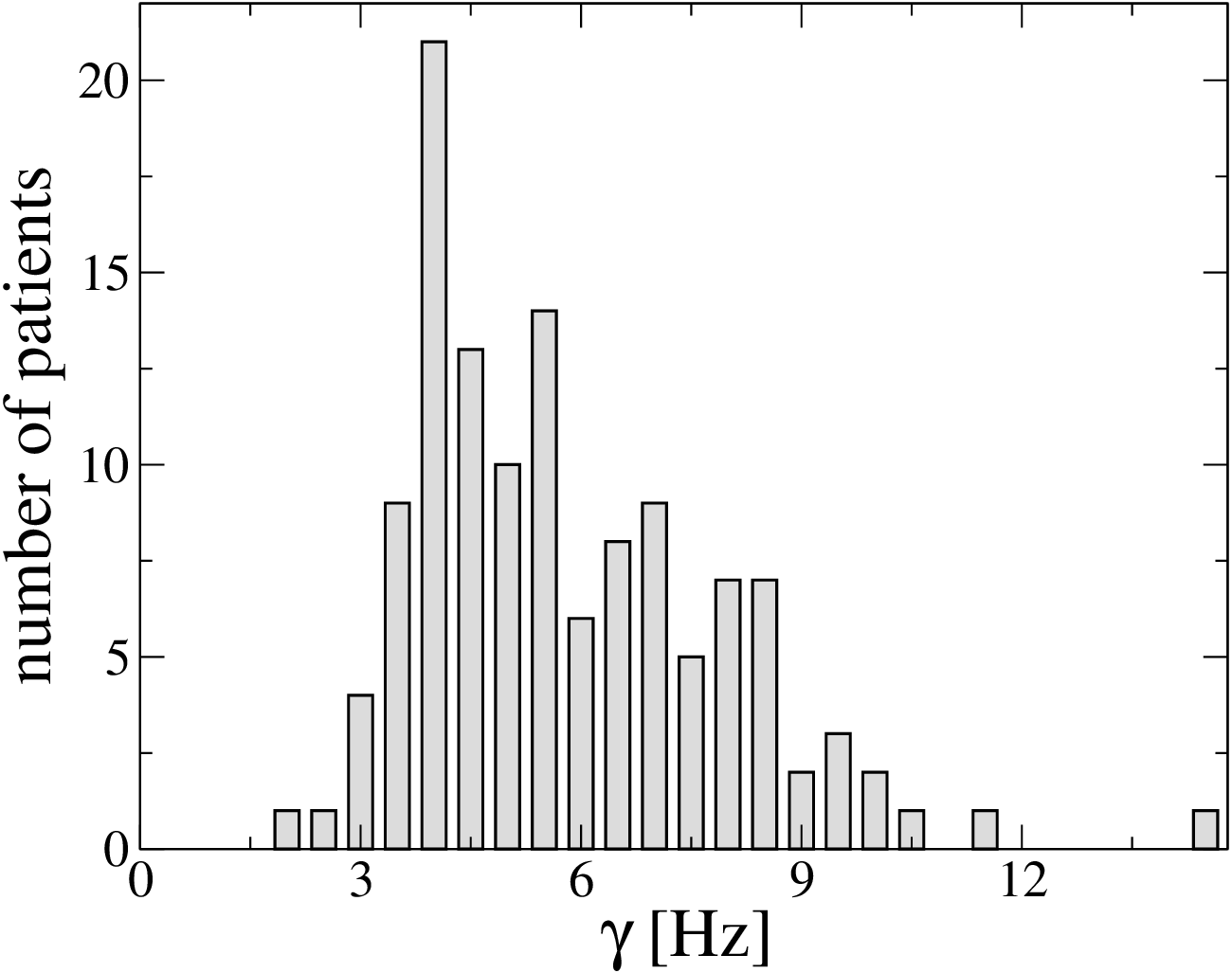}\\[0.5ex]
\caption{Histogram of $\gamma$ values 
  with mean 5.4~Hz and standard deviation 2.1~Hz for all 125
  patients.}
\label{fig:gamma}
\end{figure}

The rough method to identify the tail yields approximate $\gamma$
values in the range 3-12~Hz. However, to determine the parameters
$p_\infty$ and $\gamma$ for each patient more precisely, a ``best
fitting'' interval $[\tau_1,\tau_2]$ is calculated for each patient,
in which the Kolmogorov-Smirnov deviation is smallest. In addition to
the requirement that the tail region should encompass at least 2\% of
the total number of RR intervals, we vary both $\tau_1$ and $\tau_2$
under the constraints {\it (i)} $\tau_2-\tau_1\ge0.3\,{\rm s}$ and
{\it (ii)} $\tau_1>0.7\,{\rm s}$. The first constraint {\it (i)}
guarantees that, in the case of small $\gamma$, the width of the
fitting region covers about one decay time $1/\gamma$. In the case of
large $\gamma$, it ensures that the fitting interval covers several
$1/\gamma$ so that it is not influenced by short-time fluctuations.
The second constraint {\it (ii)} in combination with {\it (i)} ensures
that the fitting region lies right to the maximum of $p(\tau)$. In the
best fitting region, 125 of the 130 AF patients (96.2\%) pass the
Kolmogorov-Smirnov test with significance level 10\% and these are
taken into account in the following analysis. The fitting region has
a mean width of 0.42~s and on average comprises 8.6\% of the RR
intervals. The histogram of $\gamma$ values for these patients is
displayed in Fig.~\ref{fig:gamma}. The values lie in the range 2-12~Hz
and have a mean 5.4~Hz.

While almost all $p(\tau)$ of the AF patients show the exponential
tail, the shapes of the distributions differ outside the tail region.
The shapes can be grouped into classes according to the number of
local maxima in $p(\tau)$ (unimodal or multimodal distributions). The
physio\-logical origin of multimodal distributions is still under
investigation and mostly associated with the presence of different
conduction pathways through the AV node
\cite{Rokas:2001,Hatzidou:2002,Weismuller:2001}.

\begin{figure}[h!]
\vspace*{1em}
  \includegraphics[width=10cm,clip=,]{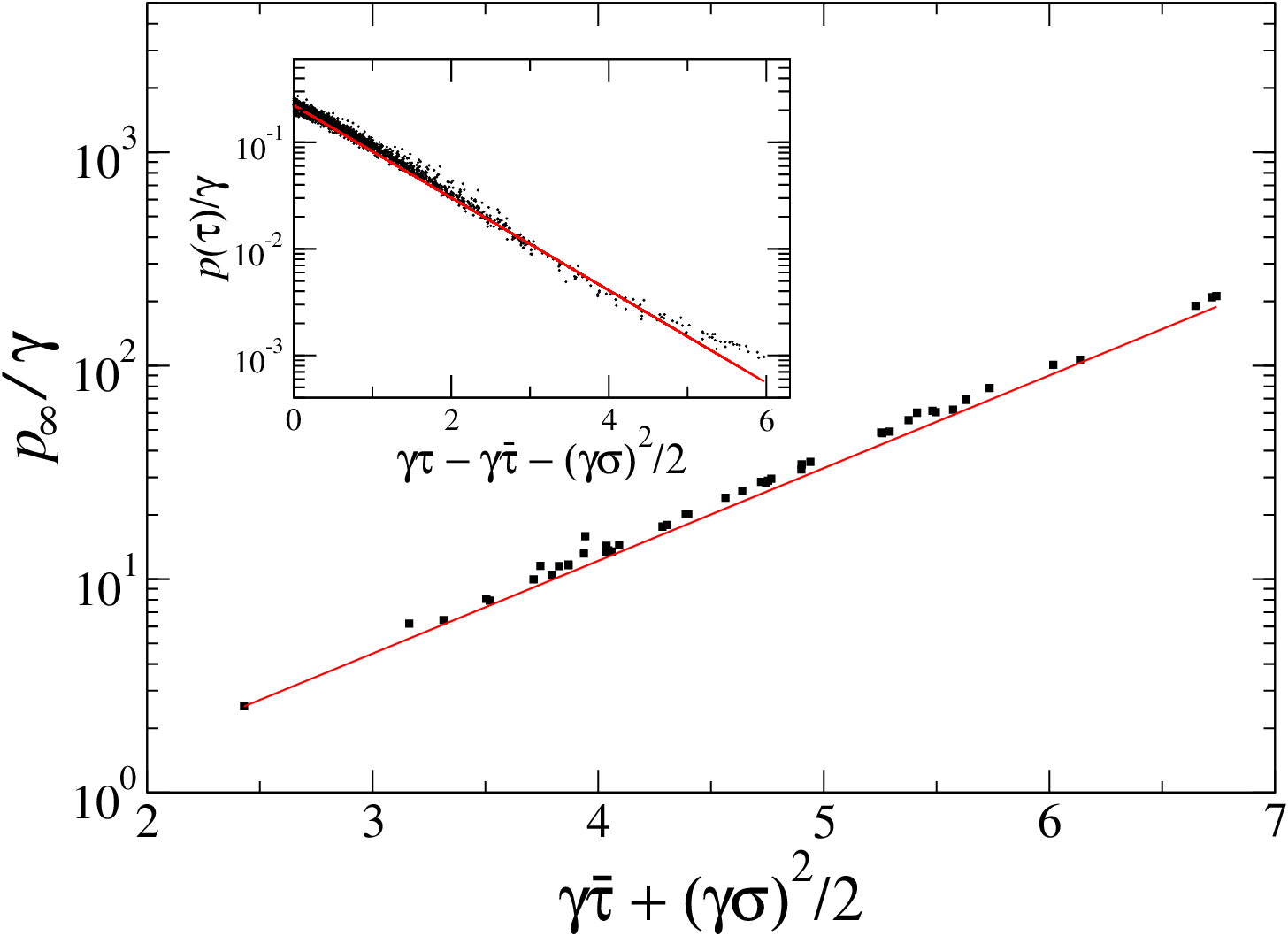}\\[0.5ex]
\caption{Semi-logarithmic plot of $p_\infty/\gamma$ for the subgroup of 
  48 patients as a function of $\gamma\bar\tau+\gamma^2\sigma^2/2$.
  The straight line marks the theoretical behaviour according to
  eq.~(\ref{eq:pinf}).  The inset shows the rescaled probability density
  for this group of patients.} 
\label{fig:pinf}
\end{figure}

A full account for the unimodal distributions can be obtained by
assuming that $\tau$ results from a superposition of two statistically
independent times, $\tau=\tau'+\eta$, where $\tau'$ is drawn from a
Gaussian distribution $\psi(\tau')$ with mean $\tau_{\rm G}$ and
variance $\sigma_{\rm G}^2$, and $\eta$ is drawn from an exponential
distribution $\phi(\eta)=\gamma\exp(-\gamma\eta)$ \cite{decomp-comm}.
In order for the times $\tau'$ to be positive one could formally
introduce a truncated Gaussian, but because it turns out that
$\psi(\tau')$ is sufficiently sharply peaked at $\tau_{\rm G}$, we
can ignore the truncation. Then we obtain
\begin{eqnarray}
p(\tau)&=&\int_{-\infty}^\tau d\tau'\,\psi(\tau')\,\phi(\tau-\tau')\nonumber\\
&=&\frac{\gamma\exp(\gamma\tau_{\rm G}+\gamma^2\sigma_{\rm G}^2/2)}{2}\,
{\rm erfc}\left(\frac{\tau_{\rm G}+\gamma\sigma_{\rm G}^2-\tau}
                      {\sqrt{2}\sigma_{\rm G}}\right)
\exp(-\gamma\tau)
\label{eq:pfit}
\end{eqnarray}
with mean $\bar\tau=\tau_{\rm G}+\gamma^{-1}$ and variance
$\sigma^2=\sigma_{\rm G}^2+\gamma^{-2}$. Hence, given the $\gamma$
values as determined above, fits to eq.~(\ref{eq:pfit}) can be easily
performed by just calculating $\bar\tau$ and $\sigma^2$. These fits
are in good agreement for all 48 AF patients with a unimodal
distribution. For the representative example in
Figs.~\ref{fig:density}a,b. this is demonstrated by the solid lines.
For large RR intervals the distribution follows eq.~(\ref{eq:decay})
with
\begin{equation}
p_\infty=\gamma\exp\left(\gamma\tau_{\rm G}+\frac{\gamma^2\sigma_{\rm
    G}^2}{2}\right)=
\gamma\exp\left(\gamma\bar\tau+\frac{\gamma^2\sigma^2}{2}-\frac{3}{2}\right)
\label{eq:pinf}
\end{equation}
As shown in Fig.~\ref{fig:pinf} the $p_\infty$ determined by
the linear regression follow closely the theoretical curve predicted
by eq.~(\ref{eq:pinf}). This allows us to scale the exponential tails
for these AF patients onto a common master curve, see the inset of
Fig.~\ref{fig:pinf}. The $\tau_{\rm G}$ and $\sigma_{\rm G}$
lie in the ranges $0.3-0.85$~s\ and $0.04-0.17$~s\,
respectively. 

The decay rate $\gamma$ can be viewed as a new parameter for a further
classification of AF. As such, it may be subjected to diurnal
variations as, for example, the fibrillation rate
\cite{Meurling/etal:2001}. To test this, we next study correlations
between $\gamma$ and $\tau_{\rm G}$, which can be expected to be
larger during night than day time. The sequences of RR intervals are
split into segments of $2^{14}$ intervals (corresponding to typically 4.5
hours). For good temporal resolution, the analysis is performed on
overlapping segments shifted by $2^{12}$ beats (corresponding to
typically 1 hour). In each segment $i$, $\gamma_i$ is determined
from the RR interval distribution belonging to this segment as
described above, and $\tau_{{\rm G},i}=\bar\tau_i-\gamma^{-1}_i$ 
is calculated from the the mean $\bar\tau_i$ in the segment.

Figure~\ref{fig:correlation}a shows the results for a representative
patient. Clearly, time periods with larger $\tau_{\rm G}$ correspond
to periods of smaller $\gamma$ and vice versa. To quantify this
behaviour we calculated the cross correlation coefficient $C$ for all
AF patients. The histogram of $C$ values in
Fig.~\ref{fig:correlation}b reflects pronounced anti-correlations. The
mean correlation coefficient is $\bar C=-0.86$.

\section{Relation to properties of the power spectrum}
\label{sec:power-spectrum}

\begin{figure}[h!]
\vspace*{2em}
\includegraphics[width=12cm,clip=,]{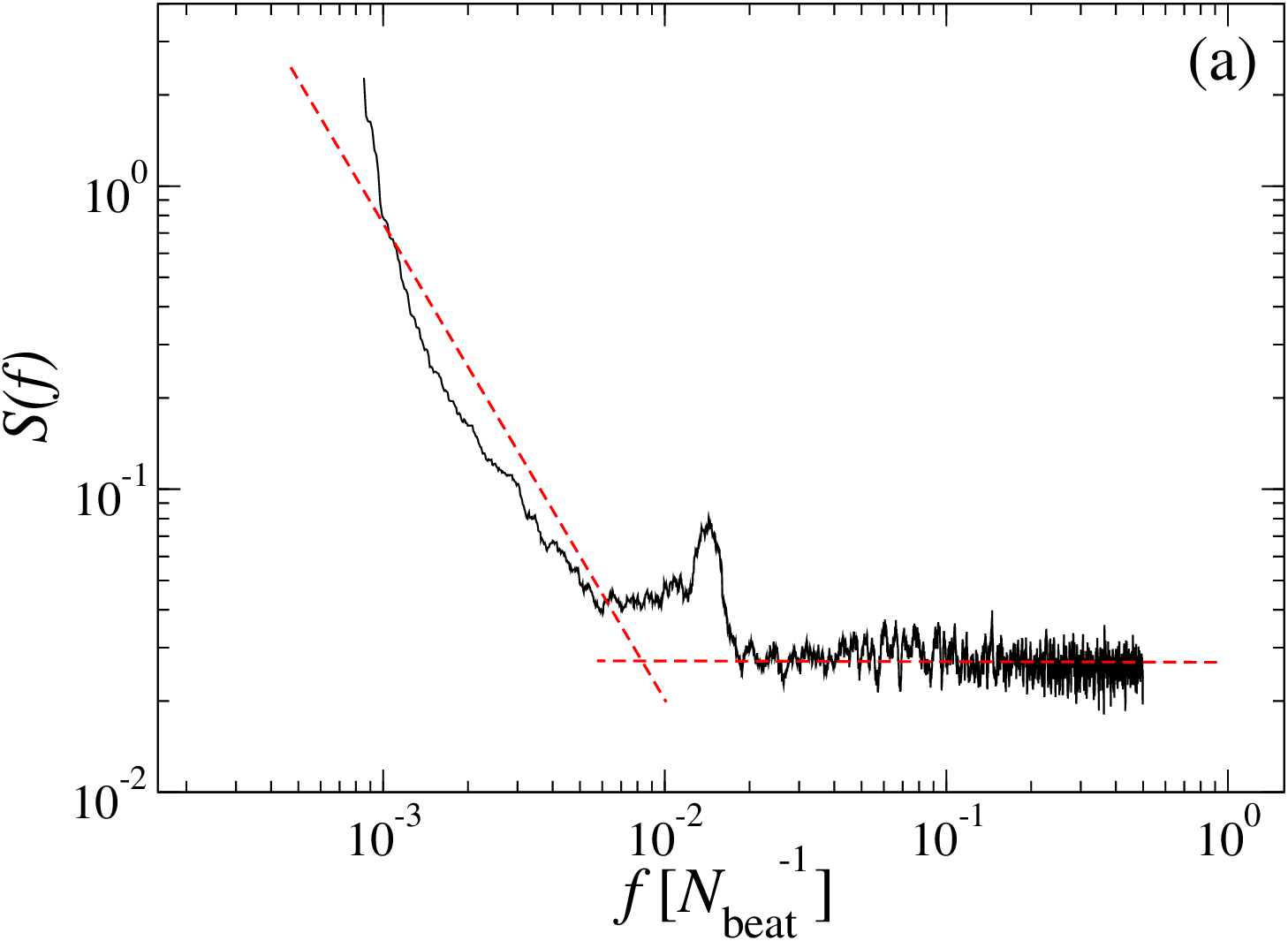}\\[3ex]
\hspace*{0.4em}
\includegraphics[width=11.8cm,clip=,]{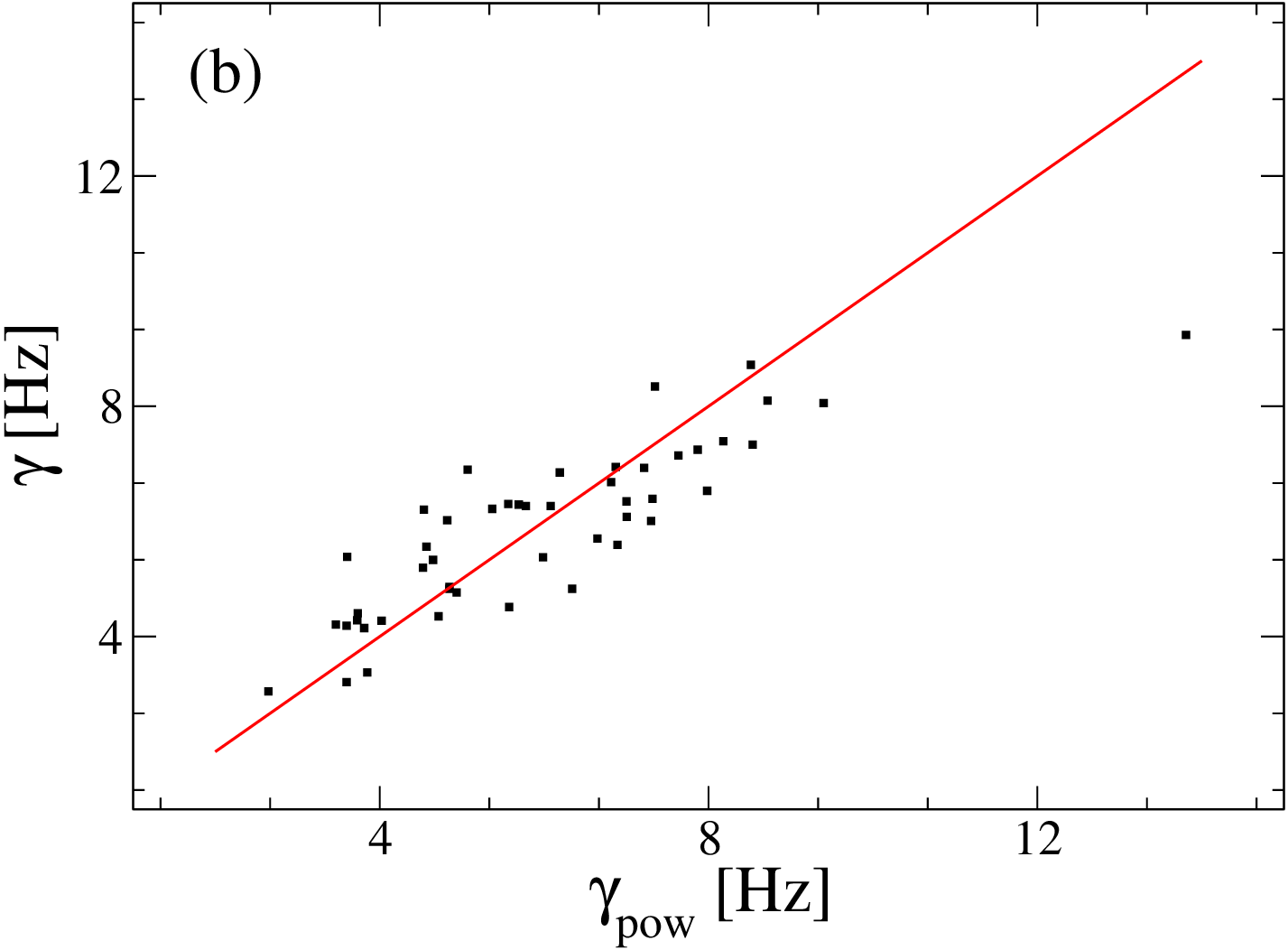}\\[0.5ex]
\caption{{\it (a)} Power spectrum of RR intervals $\tau$ 
  for one representative AF patient on a double-logarithmic scale. The
  spectrum exhibits a distinct crossover at a frequency $f_{\rm
    crit}\simeq (100\;{\rm beats})^{-1}$. For frequencies $f<f_{\rm
    crit}$ the spectrums exhibit the typical $1/f$ behaviour and a
  white noise characteristic for $f>f_{\rm crit}$. {\it (b)} $\gamma$
  vs.  $\gamma_{\rm pow}$ for the group of 48 patients with unimodal
  $p(\tau)$. The red line marks the theoretical relation
  $\gamma=\gamma_{\rm pow}$.}
\label{fig:Power_Gamma}
\end{figure}

Properties of the power spectrum $S(f)$ of the RR interbeat intervals
\cite{power-comm} during AF have been studied
previously by some of the authors and coworkers
\cite{Hayano/etal:1997,Heinrichs/etal:2004}. For low frequencies
$S(f)$ exhibits a $1/f$ type behaviour as known also for healthy
persons. At higher frequencies, however, striking differences between
the power spectra of healthy persons and AF patients are found (see
Fig.~\ref{fig:Power_Gamma}a for a representative example). While the
$1/f$ type behaviour continues up to the highest frequencies for the
healthy persons, there is a crossover to an extended part with a
behaviour close to white noise at high frequencies for the AF patients.
The question arises if the two characteristic features, i.e. the
exponential tail in the distribution of RR intervals and the white
noise part in the power spectrum, are interrelated due to a generic
underlying mechanism.

Motivated by the successful decomposition of the RR intervals $\tau$
into two contributions $\tau'$ and $\eta$, we suppose the
exponentially distributed times $\eta$ to be uncorrelated and hence
associated with the white noise part in $S(f)$. By contrast the time
$\tau'$ should reflect a strongly correlated process associated with
the $1/f$ part. Accordingly we write $S(f)=S_{\tau'}(f)+S_{\eta}(f)$,
where $S_{\tau'}(f)\sim S_0f^{-\alpha}$ and $S_\eta(f)\simeq S_\infty$
are the power spectra of the two contributions. Then the crossover
frequency is given by $f_\times=(S_0/S_\infty)^{1/\alpha}$.  The
strength $S_\infty$ of the white noise should be given by the decay
rate $\gamma$ (since $S_\infty=S_\eta(f)=S_\eta(f\to0)$, and the zero-frequency
limit of a power spectrum equals the variance),
\begin{equation}
S_\infty=1/\gamma^2\,.
\label{eq:sinf} 
\end{equation}

\begin{figure}[h!]
\vspace*{2em}
\includegraphics[width=12cm,clip=,]{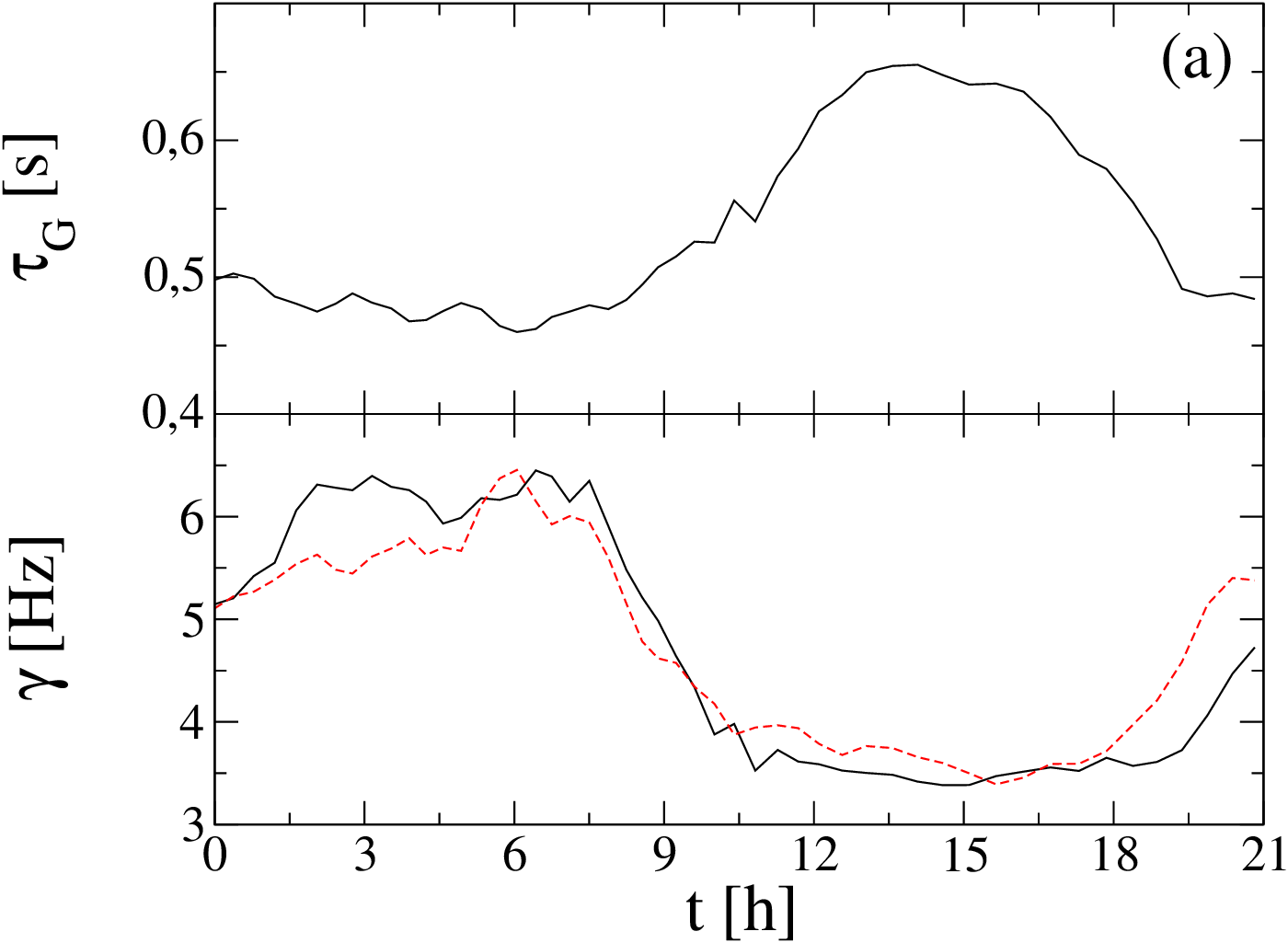}\\[3ex]
\hspace*{1.5em}
\includegraphics[width=11.2cm,clip=,]{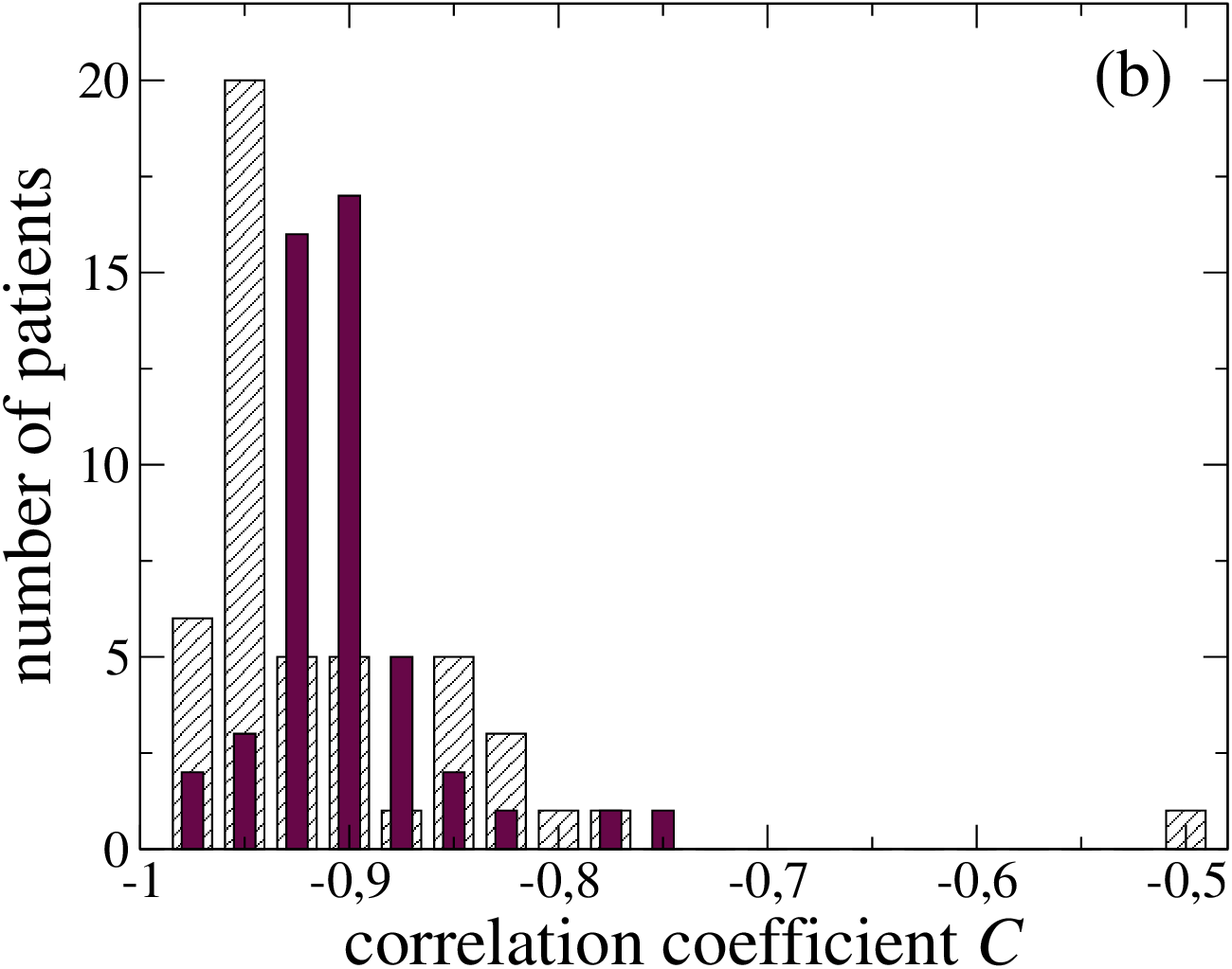}\\[0.5ex]
\caption{{\it (a)} Time dependence of
  $\tau_{\rm G}$ and decay rate $\gamma$ in segments of $2^{14}$ beats
  during AF (solid line: $\gamma$ calculated from the RR interval
  distribution; dashed line: $\gamma=\gamma_{\rm pow}$ calculated from
  the power spectrum, see Sec.~\ref{sec:power-spectrum}). The segments
  were shifted by $2^{12}$ beats, and $\tau_{{\rm G},i}$ and
  $\gamma_i$ are assigned to the starting point of each segment $i$.
  {\it (b)} Histogram of cross-correlation coefficients
  $C=\sum_{i=1}^N(\tau_{{\rm
      G},i}-\tilde\tau)(\gamma_i-\tilde\gamma)/(N\Delta_\gamma
  \Delta_\tau)$, where $N$ is the number of segments,
  $\tilde\tau=\sum_{i=1}^N\tau_{{\rm G},i}/N$,
  $\tilde\gamma=\sum_{i=1}^N\gamma_i/N$, and $\Delta_\tau$,
  $\Delta_\gamma$ are the corresponding standard deviations (full
  bars: $\gamma_i$ calculated from the RR interval distribution;
  hatched bars: $\gamma_i$ calculated from the power spectrum, see
  Sec.~\ref{sec:power-spectrum}).}
\label{fig:correlation}
\end{figure}

Using (\ref{eq:sinf}), we determined $\gamma_{\rm pow}\equiv
S_\infty^{-1/2}$ by averaging $S_{\eta}(f)$ over the modes belonging
to the 100 largest frequencies \cite{power-comm}. The result is shown
in Fig.~\ref{fig:Power_Gamma}b in comparison with the decay rate
$\gamma$ obtained from $p(\tau)$. A fair agreement is found for all 48
AF patients. Also, $\gamma_{\rm pow}$ follows closely the diurnal
variations of $\gamma$. This can be seen by the dashed line
and the hatched bars in Figs.~\ref{fig:correlation}b and
\ref{fig:correlation}c, respectively. Nevertheless, when doing the
time-resolved analysis, the $\gamma$ obtained from the RR
interval distribution and the power spectrum are not as close as in
Fig.~\ref{fig:correlation}b. This is due to the fact that a precise
determination of $\gamma$ from the RR interval distribution
becomes rather difficult for a comparatively small number of RR
intervals. By contrast, the white noise part in the power spectrum can
be clearly identified already for segments comprising $2^{10}$ beats
only. As a consequence, the power spectrum should be preferred for the
determination of the new parameter $\gamma$ in medical applications.

\section{Conclusions}

In summary we have studied the distribution and correlation properties
of heart beat intervals based on long-term (24h) ECG recordings during
AF. A characteristic exponential tail in the distribution of heart
beat intervals was found, which is absent in healthy controls and can
therefore be used as a novel way to characterise AF. We quantify the
tail with the decay rate $\gamma$ and find values in the range
3--12~Hz. Surprisingly, these $\gamma$ values can also be determined
from the white noise part of the power spectrum. The findings can be
described by decomposing the RR intervals into two statistically
independent times, where the first one belongs to a correlated process
with the typical $1/f$ noise characteristics (as found also for
healthy subjects), while the second one belongs to an uncorrelated
process. The uncorrelated process is responsible for the occurrence of
the exponential distribution and the white noise. Over the day the
$\gamma$ values vary slowly and these variations are anticorrelated
with the mean interval belonging to the correlated process. While the
time-resolved analysis is rather difficult to perform based on the
distribution function, it is straightforwardly carried out based on
the power spectrum.

It is remarkable that the range of $\gamma$ values as well as its
diurnal variations are comparable to the range and diurnal variations
of the fibrillation rates typically found in medical studies
\cite{Gaita/etal:1998,Hobbs/etal:2000,Meurling/etal:2001}. While it is
currently unclear if this is a mere accident or if there exists a
deeper connection, the potential of the new parameter $\gamma$ for the
clinical diagnosis should be evaluated. This parameter is readily
available from the surface ECG and accordingly easier to obtain than
data from intra-atrial recordings. Similar to the fibrillation rate,
$\gamma$ may provide a useful measure to estimate the outcome of a
cardioversion or of pharmaceutical treatments. Among others, it could
also help to predict the risk that patients with paroxysmal AF develop
chronical AF. We hope that our findings will stimulate further
research in this direction.

Apart from this issue of practical importance, the results require an
explanation based on physiological mechanisms connected with AF. For
such explanation both the statistics of atrial impulses and the
transduction of them through the atrio-ventricular node are expected to
play a key role.  On may assume that the observed statistics directly
reflect statistical features of the intra-atrial beats. However,
typical recordings in the right atria show a rather narrow
distribution of the intervals between successive atrial impulses and
therefore do not support this idea. On the other hand, it has been
suggested that blocked atrial impulses can prolongate the (absolute or
relative) refractory period of the atrio-ventricular node, and this
process could be a possible reason for the peculiar statistics of the
long heart beat intervals during AF. Further studies, however, are
needed to clarify this possibility.

\newpage

\end{article}

\begin{thebibliography}{99}
\bibitem{decomp-comm} For multimodal distributions one can adopt an
  analogous procedure but should take into account that $\tau'$ is
  drawn from a convolution of several Gaussians.
\bibitem{power-comm} The power spectrum of the sequence of RR
  interbeat intervals $\tau_n$, $n=0,\ldots,N-1$ is calculated by
  $S(f_m)=|\hat\tau(f_m)|^2/N$ for $f_m=m/N$, $m=-N/2,\ldots,N/2-1$,
  where $\hat\tau(f_m)=\sum_{n=0}^{N-1}(\tau_n-\bar\tau)\,\exp(-2\pi
  if_mn)$.
\end{thebibliography}
\end{document}